\documentstyle[twocolumn,psbox,myletter]{jpsj}

\def\runtitle{Spin Dependence of the Critical Concentration for the N\'{e}el State of 2D Impure Heisenberg Antiferromagnets }
\def\runauthor{Chitoshi {\sc Yasuda} and Akihide {\sc Oguchi}}

\title
{
Spin Dependence of the Critical Concentration for the N\'{e}el State of 2D Impure Heisenberg Antiferromagnets
}

\author
{ 
Chitoshi {\sc Yasuda}\footnote{Present address : Institute for Solid State Physics, The University of Tokyo, 7-22-1 Roppongi, Minato-ku, Tokyo 106-8666.}  and Akihide {\sc Oguchi}
}

\inst
{
Department of Physics,
Faculty of Science and Technology, Science University of Tokyo, Noda, Chiba 278-8510
}

\recdate
{
June 7, 1999
}

\abst
{
The spin-wave theory and the coherent potential approximation are applied to a spin $S$ Heisenberg antiferromagnet with nonmagnetic impurities on square lattice. The impurity effects are taken into account by substituting $S(1-x)$ for $S$ and using the coherent potential approximation to the exchange interaction, where $x$ is the impurity concentration. At $T=0$ for $S=1/2$ the critical impurity concentration $x_{\rm c}$ of the N\'{e}el state is 0.303 and the percolation threshold $x_{\rm p}$ is 0.500. The ground state in $x_{\rm c}<x<x_{\rm p}$ is the disordered state with the spin gap. For $S \geq 1$ the long range N\'{e}el order vanishes at $x_{\rm p}=0.500$. These results explain qualitatively the experimental results of ${\rm La}_{2}{\rm Cu}_{1-x}{\rm Mg}_{x}{\rm O}_{4}~(S=1/2)$ and ${\rm K}_{2}{\rm Mn}_{1-x}{\rm Mg}_{x}{\rm F}_{4}~(S=5/2)$. The difference of $x_{\rm c}$ between these materials is caused by the decrease in the magnitude of the effective spin with impurity doping. The spin gap is expected to be observed for ${\rm La}_{2}{\rm Cu}_{1-x}{\rm Mg}_{x}{\rm O}_{4}$ in $x_{\rm c}<x<x_{\rm p}$ at low temperatures. 
}

\kword
{
${\rm La}_{2}{\rm Cu}_{1-x}{\rm Mg}_{x}{\rm O}_{4}$, ${\rm K}_{2}{\rm Mn}_{1-x}{\rm Mg}_{x}{\rm F}_{4}$, nonmagnetic impurity, critical concentration, percolation threshold, spin gap, order-disorder transition, spin-wave theory, modified spin-wave theory, CPA
}

\begin{document}
\sloppy
\maketitle

\section{Introduction}

Two dimensional magnet with nonmagnetic impurities shows interesting properties in relation to the magnetic order and the percolation.[1-12] The N\'{e}el temperature of an antiferromagnet is depressed with nonmagnetic impurity doping. The critical impurity concentration for the two dimensional (2D) Ising antiferromagnet, e.g., ${\rm K}_{2}{\rm Co}_{1-x}{\rm Mg}_{x}{\rm F}_{4}$ agrees well with the percolation threshold 0.41. The critical concentration for the 2D Heisenberg antiferromagnet, e.g., ${\rm K}_{2}{\rm Mn}_{1-x}{\rm Mg}_{x}{\rm F}_{4}$ also agrees with the percolation threshold 0.41. The agreement between the critical concentration and the percolation threshold can be understood by the fact that the long range order cannot exist if the percolation path does not connect to infinity. In compared with above two materials, by extrapolating the experimental data it is found that the critical concentration of ${\rm La}_{2}{\rm Cu}_{1-x}{\rm Mg}_{x}{\rm O}_{4}$, expressed by the antiferromagnetic Heisenberg model with nonmagnetic impurities on square lattice, is not the percolation threshold 0.41 but 0.2-0.25.[1] We consider this difference may be caused by the magnitude of the spin, $S$, of the host material, i.e., ${\rm Cu}^{2+}(S=1/2)$ of ${\rm La}_{2}{\rm Cu O}_{4}$ and ${\rm Mn}^{2+}(S=5/2)$ of  ${\rm K}_{2}{\rm Mn F}_{4}$.

If the spin operator of the Heisenberg model on square lattice is approximated by the classical vector, we can obtain the spin dependence of the critical concentration of the N\'{e}el state. In this approximation, the N\'{e}el state energy is $-4NJS^{2}$ and the singlet dimer state energy is $-NJS(S+1)$, where $N$ is the lattice number and $J$ is the exchange integral. If the nonmagnetic impurity is doped and we adopt the coherent potential approximation, then $J$ would be replaced by the coherent exchange integral $J_{\rm p}$. Comparing these energies, we find that the N\'{e}el state is stable when $S>1/3$. However, spin $S$ is allowed only to be an integer or a half odd number. This $1/3(\equiv S^{\rm cri})$ may be regarded as the effective spin $S^{\rm eff}$ averaged over all impurity configurations. As the spin $S^{\rm eff}$ is reduced to $S^{\rm cri}$ by the impurity doping, the critical impurity concentration $x_{\rm c}$ satisfies the relation,  $S(1-x_{\rm c})=S^{\rm cri}=\frac{1}{3}$. At $S=1/2$ this critical concentration $x_{\rm c}$ is about 0.3 from $\frac{1}{2}(1-x_{\rm c})=\frac{1}{3}$. Thus if the nonmagnetic impurity doping reduces the effective spin, the singlet state can be stable when $S^{\rm eff}<1/3$ and $x>0.3$. On the other hand, at $S=5/2$ the critical concentration $x_{\rm c}$ is about 0.87 and is larger than the percolation threshold 0.41. Thus the antiferromagnetic order for $S=5/2$ persists up to the percolation threshold.

The critical concentration for the antiferromagnetic Heisenberg model with $S=1/2$ on square lattice has been investigated by the present authors using the Kondo-Yamaji's Green's function method with nonmagnetic impurities.[13] The obtained results show that the critical concentration is not equal to the percolation threshold and the disordered state which is characterized by the finite spin gap and the exponentially decaying correlation function appears in the range of $x_{\rm c}<x<x_{\rm p}$. However, the generalization of the Kondo-Yamaji's method to the general spin $S$ is difficult, because we cannot self-consistently determine a new decoupling parameter. On the other hand, the spin-wave theory can be applied to any spin systems. Therefore, in this paper we study by using the spin-wave theory and the modified spin-wave theory.[14]

The present paper is organized as follows: in {\S} 2, we formulate the spin-wave theory for 
a 2D Heisenberg antiferromagnet with nonmagnetic impurities and obtain the critical concentration and the percolation threshold for any spin $S$ at $T=0$; in {\S} 3, the modified spin-wave theory is applied in the range between the critical concentration and the percolation threshold. Section 4 is devoted to summary and discussion.

\section{Spin-wave theory of a 2D Heisenberg antiferromagnet with nonmagnetic impurities}

We consider the antiferromagnetic ($J>0$) Heisenberg model with nonmagnetic impurities on  square lattice:
\begin{equation}
   \label{ham}
   H^{\rm c} =  2J\sum_{<i,j>} \sigma_i \sigma_j {\mib S}_i \cdot {\mib S}_j \ ,
\end{equation}
where $\sum_{<i,j>}$ denotes the summation over all nearest neighbor sites. The magnetic occupation operator $\sigma_i$ are defined by 
\begin{equation}
       \sigma_i = \left\{
                  \begin{array}{@{\,}ll}
                     1 & \mbox{if \ the \ site \  {\it i} \ is \  occupied \ by \ a \ spin} \ , \\
                     0 & \mbox{otherwise} \ .
                  \end{array}
                  \right. \
\end{equation}
The subscript c stands for a given spin configuration. Assuming the lattice to be bipartite, we use $i,l,l'\in A$ sublattice and $j,m,m'\in B$ sublattice. We introduce the antiferromagnetic Holstein-Primakoff transformation for the impure case:
\begin{eqnarray}
 \label{HPtrans1}
 && \left\{
    \begin{array}{@{\,}l}
     \sigma_{l}S_{l}^{-}=\sigma_{l}a_{l}^{\dagger}\sqrt{2S'-\sigma_{l}a_{l}^{\dagger} a_{l}} \\
     \sigma_{l}S_{l}^{+}=\sqrt{2S'-\sigma_{l}a_{l}^{\dagger} a_{l}}~\sigma_{l}a_{l} ~~ , \\  \sigma_{l}S_{l}^{\rm z}=S'-\sigma_{l}a_{l}^{\dagger} a_{l} 

    \end{array}
    \right. \\ 
 \label{HPtrans2}
 && \left\{
    \begin{array}{@{\,}l}
     \sigma_{m}S_{m}^{-}=\sqrt{2S'-\sigma_{m}b_{m}^{\dagger} b_{m}}~\sigma_{m}b_{m} \\
     \sigma_{m}S_{m}^{+}=\sigma_{m}b_{m}^{\dagger}\sqrt{2S'-\sigma_{m}b_{m}^{\dagger} b_{m}}~~ , \\     \sigma_{m}S_{m}^{\rm z}=-S'+\sigma_{m}b_{m}^{\dagger} b_{m} 

    \end{array}
    \right. 
\end{eqnarray}
where $a_{l}$ and $b_{m}$ are bose operators and satisfy the relations $[a_{l},a_{l'}^{\dagger}]=\delta_{ll'}$ and $[b_{m},b_{m'}^{\dagger}]=\delta_{mm'}$. Here $S' \equiv S \langle \sigma_{l} \rangle_{\rm av} = S(1-x)$ and $x$ is the impurity concentration. The bracket  $\langle \cdots \rangle_{\rm av}$ means the average over all configurations. In order to introduce the decrease in the number of spin with impurity doping into our theory, we replace $S \sigma_{l}$ with $S \langle \sigma_{l} \rangle_{\rm av}$. When the spin $S$ is substituted by the nonmagnetic impurity, bonds with the exchange integral $J$ are randomly configured. The randomness of both the on-site spin and the exchange interaction must be considered. Since the Heisenberg model has only the interaction between spins, the randomness of the exchange integral can be treated in the frame of the coherent potential approximation (CPA) with the Tahir-Kheli's two-site approximation. And also the on-site randomness can be expressed by replacing $S$ with $S'$. Rewriting the Hamiltonian (\ref{ham}) by eqs. (\ref{HPtrans1}) and (\ref{HPtrans2}) and ignoring terms more than the zeroth order of $1/S'$, we obtain
\begin{equation}
   \label{ham1}
   H^{\rm c} = 2S'\sum_{<i,j>} J\sigma_i \sigma_j (a_{i}^{\dagger}a_{i}+b_{j}^{\dagger}b_{j}+a_{i}b_{j}+a_{i}^{\dagger}b_{j}^{\dagger}) \ .
\end{equation}
We introduce the Green's functions defined by
\begin{eqnarray}
   \label{green}
    &&  G_{ll'}^{\rm c}(t-t')  \equiv  \langle\langle \sigma_l a_{l}(t);\sigma_{l'} a_{l'}^{\dagger} (t') \rangle\rangle^{\rm c} \ , \\
    && F_{ml'}^{\rm c}(t-t')  \equiv  \langle\langle \sigma_m b_m^{\dagger} (t);\sigma_{l'} a_{l'}^{\dagger}(t') \rangle\rangle^{\rm c} \ ,          
\end{eqnarray}
where $\langle \langle A ; B \rangle \rangle^{\rm c} \equiv -{\rm i}\theta (t-t')\langle [A,B] \rangle^{\rm c}$ and $\langle A \rangle^{\rm c} \equiv {\rm Tr} A{\rm e}^{-\beta H^{\rm c}}/{\rm Tr} {\rm e}^{-\beta H^{\rm c}}$. Energy Fourier transforms of equations of motion are expressed by
\begin{eqnarray}
   \label{motion1}
 &&  \omega G_{ll'}^{\rm c}(\omega) = \sigma_{l}\delta_{ll'}
      + S'\sum_{\delta} J\sigma_{l}\sigma_{l+\delta}  \nonumber \\
 && \hspace*{2.7cm} {} \times  (G_{ll'}^{\rm c}(\omega)+F_{l+\delta,l'}^{\rm c}(\omega)) \ ,~~  \\
   \label{motion2}
 &&    \omega F_{ml'}^{\rm c}(\omega) = 
      -S'\sum_{\delta} J\sigma_{m}\sigma_{m+\delta} \nonumber \\
 &&  \hspace*{2.7cm} {} \times  (G_{m+\delta,l'}^{\rm c}(\omega)+F_{ml'}^{\rm c}(\omega)) \ ,~~ 
\end{eqnarray}
where $\delta$'s are z vectors to nearest neighbors and $G_{ll'}^{\rm c}(\omega)$ and $F_{ml'}^{\rm c}(\omega)$ are Fourier transforms of $G_{ll'}^{\rm c}(t-t')$ and $F_{ml'}^{\rm c}(t-t')$, respectively. 

In order to treat the nonmagnetic impurity effect of the exchange integral, we introduce the coherent exchange integral $J_{\rm p}(\omega)$.[15] Adding the same terms to both sides of eqs. (\ref{motion1}) and (\ref{motion2}), we can write eqs. (\ref{motion1}) and (\ref{motion2}) as a matrix expression:
\begin{eqnarray}
    \label{matrix}
&&  \hspace*{1.5cm} \Gamma^{-1} g^{\rm c} = \Lambda^{\rm c} + V^{\rm c} g^{\rm c} \ , \\
    \Gamma^{-1} &\equiv& 
     \left(
       \begin{array}{cc}
          (\Gamma^{-1})_{li} & (\Gamma^{-1})_{lj} \\
          (\Gamma^{-1})_{mi} & (\Gamma^{-1})_{mj}
       \end{array}
     \right) \nonumber \\ 
  &=& 
     \left(
    \begin{array}{cc}
       [\omega-J_{\rm p}(\omega)S'z] \delta_{il} & 
       -J_{\rm p}(\omega)S' \sum_{\delta} \delta_{j,l+\delta} \\
       J_{\rm p}(\omega)S' \sum_{\delta}\delta_{i,m+\delta} &
       [\omega+J_{\rm p}(\omega)S'z] \delta_{jm} 
    \end{array}
    \right) \ , \nonumber \\
    g^{\rm c} &\equiv& 
     \left(
       \begin{array}{c}
            G_{ll'}^{\rm c}(\omega) \\ F_{ml'}^{\rm c}(\omega) 
       \end{array}
     \right)  \ , \hspace{0.5cm} \Lambda^{\rm c} \equiv 
     \left(
       \begin{array}{c}
             \sigma_{l}\delta_{ll'} \\ 0
       \end{array}
     \right) \ , \nonumber \\
   V^{\rm c} &\equiv&
     \left(
       \begin{array}{cc}
          V_{li}^{\rm c} & V_{lj}^{\rm c} \\
          V_{mi}^{\rm c} & V_{mj}^{\rm c}
       \end{array}
     \right) \nonumber \\ 
     &=& 
     \left(
    \begin{array}{l}
       S'\sum_{\delta}[J\sigma_{l}\sigma_{l+\delta}-J_{\rm p}(\omega)]\delta_{il} \\
      -S'\sum_{\delta}[J\sigma_{m}\sigma_{m+\delta}-J_{\rm p}(\omega)]\delta_{i,m+\delta}     
    \end{array}
    \right.  \nonumber \\
&&  \hspace*{1.5cm} \left.
      \begin{array}{r}
          S'\sum_{\delta}[J\sigma_{l}\sigma_{l+\delta}-J_{\rm p}(\omega)]\delta_{j,l+\delta} \\
          -S'\sum_{\delta}[J\sigma_{m}\sigma_{m+\delta}-J_{\rm p}(\omega)]\delta_{jm}
      \end{array}
      \right)  \ . \nonumber      
\end{eqnarray}
Using the relation $\Gamma^{-1} \Gamma = 1$, we obtain   
\begin{eqnarray}
 && \Gamma \equiv 
     \left(
       \begin{array}{cc}
          \Gamma_{li} & \Gamma_{lj} \\
          \Gamma_{mi} & \Gamma_{mj}
       \end{array}
     \right)  \\
 &&  = 
     \left(
       \begin{array}{cc}
         \frac{2}{N}\sum_{\mibs k}\Gamma_{\mibs k}^{(00)}(\omega){\rm e}^{{\rm i}{\mibs k}\cdot({\mibs l}-{\mibs i})} & 
         \frac{2}{N}\sum_{\mibs k}\Gamma_{\mibs k}^{(01)}(\omega){\rm e}^{{\rm i}{\mibs k}\cdot({\mibs l}-{\mibs j})} \\
         \frac{2}{N}\sum_{\mibs k}\Gamma_{\mibs k}^{(10)}(\omega){\rm e}^{{\rm i}{\mibs k}\cdot({\mibs m}-{\mibs i})} &
         \frac{2}{N}\sum_{\mibs k}\Gamma_{\mibs k}^{(11)}(\omega){\rm e}^{{\rm i}{\mibs k}\cdot({\mibs m}-{\mibs j})} 
       \end{array}
     \right), \nonumber
\end{eqnarray}
with
\begin{eqnarray}
 && \Gamma_{\mibs k}^{(00)}(\omega) = -\Gamma_{\mibs k}^{(11)}(-\omega)
    = \frac{\omega +J_{\rm p}(\omega)S'z}{\omega^{2}
                       -[J_{\rm p}(\omega)S'z]^{2}(1-\gamma_{\mibs k}^{2})} \ , \hspace{20pt} \\
 && \Gamma_{\mibs k}^{(10)}(\omega) = -\Gamma_{\mibs k}^{(01)}(\omega)
    = \frac{-J_{\rm p}(\omega)S'z\gamma_{\mibs k}}{\omega^{2}
                       -[J_{\rm p}(\omega)S'z]^{2}(1-\gamma_{\mibs k}^{2})} \ , \
\end{eqnarray}
and
\begin{equation}
  \gamma_{\mibs k} \equiv z^{-1}\sum_{{\mibs \delta}}{\rm e}^{{\rm i}{\mibs k}\cdot{\mibs \delta}} \ ,
\end{equation}
where ${\mib k}$ runs over a half of the original Brillouin zone. Introducing the T-matrix through the relation:
\begin{equation}
    \label{tmat}
       V^{\rm c}g^{\rm c} = T^{\rm c}\Gamma \Lambda^{\rm c} \ ,
\end{equation}
from eqs. (\ref{matrix}) and (\ref{tmat}) we have
\begin{eqnarray}
       \label{confg}
         g^{\rm c} = \Gamma \Lambda^{\rm c} + \Gamma T^{\rm c}\Gamma \Lambda^{\rm c} \ , \\
       \label{tcv} 
         T^{\rm c} = V^{\rm c} + V^{\rm c}\Gamma T^{\rm c} \ . 
\end{eqnarray}
We use the two-site approximation to solve eq. (\ref{tcv}) for $T^{\rm c}$ and obtain
\begin{eqnarray}
&&     \left(
       \begin{array}{cc}
          T_{00}^{\rm c} & T_{01}^{\rm c} \\
          T_{10}^{\rm c} & T_{11}^{\rm c}
       \end{array}
     \right) = \frac{S'[J\sigma_{0}\sigma_{1}-J_{\rm p}(\omega)]}
               {1-S'[J\sigma_{0}\sigma_{1}-J_{\rm p}(\omega)]\Gamma_{0}} \hspace{24pt} \nonumber \\
    \label{2siteT}
&& \hspace*{4cm}  \times   
     \left(
       \begin{array}{cc}
          1 & -1 \\
          1 & -1
       \end{array}
     \right) \ ,
\end{eqnarray}
where
\begin{equation}
    \label{Gamma}
     \Gamma_{0} \equiv \frac{2}{N}\sum_{\mibs k}\frac{2J_{\rm p}(\omega)S'z(1-\gamma_{\mibs k}^{2})}
            {\omega^{2}-[J_{\rm p}(\omega)S'z]^{2}(1-\gamma_{\mibs k}^{2})} \ .
\end{equation} 
From eq. (\ref{2siteT}) the non-scattering condition $\langle T^{\rm c} \rangle_{\rm av} = 0$ of CPA leads the coherent exchange integral as
\begin{equation}
   \label{jp}
   \frac{J_{\rm p}(\omega)}{J} = 1-\frac{2x}{1+\Gamma_{\rm p}(\omega)} \ ,
\end{equation}
where
\begin{equation}
   \label{fome}
   \Gamma_{\rm p}(\omega) \equiv \frac{2}{N}\sum_{\mibs k}\frac{\omega^{2}}{\omega^{2}-[J_{\rm p}(\omega)S'z]^{2}(1-\gamma_{\mibs k}^{2})} \ .
\end{equation}
The average over impurity configurations is performed in the following way: the site 0 is occupied by a magnetic atom and the conditional averaging is taken over the two occupational states of the site 1.[15] That is, the probability that the site 1 is occupied by a magnetic atom, is $1-x$ and the probability occupied by a nonmagnetic impurity is $x$.

Averaging eq. (\ref{confg}) and assuming $\langle T^{\rm c}\Lambda^{\rm c}\rangle_{\rm av} \simeq \langle T^{\rm c}\rangle_{\rm av} \langle \Lambda^{\rm c}\rangle_{\rm av}$, we obtain the Green's function:
\begin{equation}
  \label{ongreen}
  \langle G_{ll}^{\rm c}(\omega)\rangle_{\rm av} = \frac{2}{N}\sum_{\mibs k}
       \frac{\omega \langle \sigma_{l}\rangle_{\rm av} +J_{\rm p}(\omega)S'z}
{\omega^{2}-[J_{\rm p}(\omega)S'z]^{2}(1-\gamma_{\mibs k}^{2})} \ .
\end{equation}
If the Green's functions in eqs. (\ref{motion1}) and (\ref{motion2}) are expanded in a series of $J\sigma_{i}\sigma_{j}$, the corresponding term of the second term of the numerator of eq. (\ref{ongreen}) is $J\sigma_{i}\sigma_{j}Sz$. This term does not change by multiplying $\sigma_{i}$ because $\sigma_{i}^{2}=\sigma_{i}$. In order to avoid the double count of the impurity effect, we  replace $J_{\rm p}(\omega)S'z\langle \sigma_{l} \rangle_{\rm av}$ with $J_{\rm p}(\omega)S'z$. Thus we must pay particular attention to the average of products of $\sigma_{l}$. For example, when $l'=l$, $\langle \sigma_{l} \sigma_{l'} \rangle_{\rm av} = \langle \sigma_{l} \rangle_{\rm av} \langle \sigma_{l'} \rangle_{\rm av} = \langle \sigma_{l} \rangle_{\rm av}^{2}$ must be replaced by $\langle \sigma_{l} \rangle_{\rm av}$. The correlation function is obtained from the spectral relation:
\begin{equation}
  \label{spec}
   \langle \sigma_{l}a_{l}^{\dagger}a_{l} \rangle 
         = -\frac{1}{\pi}\int_{-\infty}^{\infty}{\rm d}\omega \frac{1}{{\rm e}^{\beta \omega}-1} {\rm Im}\langle G_{ll}^{\rm c}(\omega+{\rm i}0)\rangle_{\rm av} \ .
\end{equation}
Since the spontaneous sublattice magnetization is expressed by
\begin{eqnarray}
     m \equiv \langle \sigma_{l}S_{l}^{\rm z} \rangle &=& S \langle \sigma_{l} \rangle_{\rm av} 
      - \langle \sigma_{l} a_{l}^{\dagger} a_{l} \rangle \nonumber \\ 
  \label{magn}
      &=& S(1-x) - \langle \sigma_{l} a_{l}^{\dagger} a_{l} \rangle \ ,  
\end{eqnarray}
where $\langle A \rangle \equiv \langle \langle A \rangle^{\rm c} \rangle_{\rm av}$, the impurity concentration dependence of $m$ is obtained from eqs. (\ref{jp}) - (\ref{magn}).

We consider the case $J_{\rm p}(\omega=0)$. This case corresponds to the real excitation spectrum without the damping due to the scattering of spin-waves by the disorder. We obtain $J_{\rm p}(0)$  from eqs. (\ref{jp}) and (\ref{fome}):
\begin{equation}
   \label{jpx}
    \frac{J_{\rm p}(0)}{J} = 1-2x \ .
\end{equation}  
The concentration $x_{\rm p}=0.5$, at which $J_{\rm p}$ is zero, is the percolation threshold. Substituting $J_{\rm p}(0)$ into eq. (\ref{ongreen}), we can analytically integrate eq. (\ref{spec}) and obtain the magnetization:
\begin{eqnarray}
 &&  m = (S+\frac{1}{2})(1-x) \nonumber \\
 && \hspace*{0.5cm} {} - \frac{2}{N}\sum_{\mibs k} \frac{1}{\sqrt{1-\gamma_{\mibs k}^{2}}}
             \frac{1}{2}{\rm coth}(\frac{\beta}{2} J_{\rm p}S'z\sqrt{1-\gamma_{\mibs k}^{2}}) \ . \hspace{24pt}
\end{eqnarray}
At $T=0$ we get
\begin{eqnarray}
  \label{mx0}
     m &=& (S+\frac{1}{2})(1-x) 
  - \frac{1}{2}\int \frac{{\rm d}{\mib k}}{(2\pi )^{2}}\frac{1}{\sqrt{1-\gamma_{\mibs k}^{2}}} \nonumber \\
    &=& (S+\frac{1}{2})(1-x) - 0.696602 \ .
\end{eqnarray}
The $x$ dependence of $J_{\rm p}$ and $m$ are obtained from eqs. (\ref{jpx}) and (\ref{mx0}) and are shown for $S=1/2$ in Fig. 1.
\begin{figure}[t]
  \begin{center}
    \psbox[size=0.65#1]{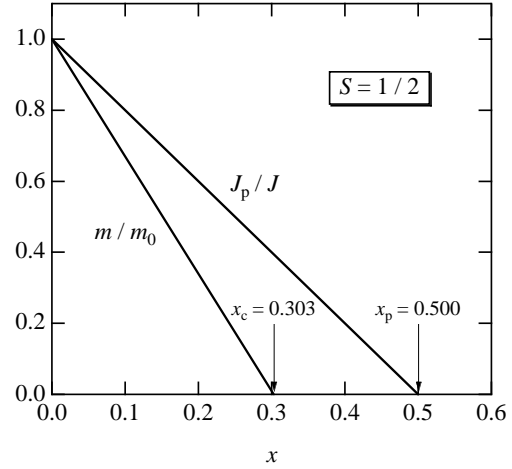}
  \end{center}
   \caption{Impurity concentration dependence of the sublattice magnetization $m$ and the coherent exchange integral $J_{\rm p}$ for $S=1/2$. The critical impurity concentration $x_{\rm c}$ is 0.303. The percolation threshold $x_{\rm p}$ is 0.500. $m_{0}$ is the sublattice magnetization for a pure system.}
\end{figure}
As $x$ is increased for $S=1/2$, both $J_{\rm p}$ and $m$ decrease and $m$ vanishes at $x_{\rm c} = 0.303$. At this concentration, however, $J_{\rm p}$ is still finite and $J_{\rm p}$ vanishes at $x_{\rm p} = 0.500$. Thus the state, with the infinite connecting percolation path and without LRO, exists in the concentration region of $x_{\rm c}<x<x_{\rm p}$. The critical concentrations of $S=1$, 3/2, 2, and 5/2 are 0.536, 0.652, 0.721 and 0.768, respectively, while the percolation thresholds of these spins are also 0.500. Since $x_{\rm c} > x_{\rm p}$ for $S \geq 1$, LRO vanishes at the percolation threshold. The spin dependence of $x_{\rm c}$ is caused by the decrease of the magnitude of the effective spin. On the other hand, $x_{\rm p}$ is independent of $S$. And so, the disagreement between $x_{\rm c}$ and $x_{\rm p}$ appears. The obtained critical concentration $x_{\rm c}=0.303$ for $S=1/2$ is disagreement with $x_{\rm c}=0.07$ obtained by the Kondo-Yamaji's method,[13] because the Kondo-Yamaji's method overestimates the spin fluctuation.

\section{Modified spin-wave theory of a 2D Heisenberg antiferromagnet with nonmagnetic impurities}

In order to investigate the state in the impurity concentration region of $x_{\rm c}<x<x_{\rm p}$, we apply the modified spin-wave theory to the 2D antiferromagnetic Heisenberg model with nonmagnetic impurities. We consider the following Hamiltonian:
\begin{equation}
   \label{ham2}
   H^{\rm c} =  2J\sum_{<i,j>} \sigma_i \sigma_j {\mib S}_i \cdot {\mib S}_j 
 + \mu (\sum_{i\in A}\sigma_i S_i^z - \sum_{j\in B}\sigma_j S_j^z) \ ,
\end{equation}
where $\mu$ is a Lagrange multiplier. The second term enforces the constraint that the total staggered magnetization is zero. We introduce the antiferromagnetic Dyson-Maleev transformation:
\begin{eqnarray}
   \label{dyson}
 &&  \left\{
    \begin{array}{@{\,}l}
     \sigma_{l} S_{l}^{\rm z}=S'-\sigma_{l} a_{l}^{\dagger} a_{l} \ ,~~ 
        \sigma_{l} S_{l}^{-}=\sigma_{l} a_{l}^{\dagger}~,~ \\
        \sigma_{l} S_{l}^{+}=(2S'-\sigma_{l} a_{l}^{\dagger} a_{l})\sigma_{l} a_{l}~,    
    \end{array}
   \right. \\   
 &&  \left\{
    \begin{array}{@{\,}l}
     \sigma_{m} S_{m}^{\rm z}=-S'+\sigma_{m} b_{m}^{\dagger} b_{m} \ ,~~ 
         \sigma_{m} S_{m}^{-}=-\sigma_{m} b_{m}~,~ \\
         \sigma_{m} S_{m}^{+}=-\sigma_{m} b_{m}^{\dagger} (2S'-\sigma_{m} b_{m}^{\dagger} b_{m})~,   
    \end{array}
   \right.    
\end{eqnarray}
and obtain equations of motion of Green's functions as 
\begin{eqnarray}
   \label{motion3}
 &&    (\omega + \mu ) G_{ll'}^{\rm c}(\omega) = \sigma_{l}\delta_{ll'} \\
 &&  ~~~~ {} + 2J\sum_{\delta} \sigma_{l}\sigma_{l+\delta} [ S' - \langle \sigma_{l+\delta}\sigma_{l+\delta}b_{l+\delta}^{\dagger}b_{l+\delta} \rangle^{\rm c} \nonumber \\ 
 &&  ~~~~ {} + \langle \sigma_{l}\sigma_{l+\delta}a_{l}b_{l+\delta} \rangle^{\rm c} ] (G_{ll'}^{\rm c}(\omega)-F_{l+\delta,l'}^{\rm c}(\omega)) \ , \nonumber \\
   \label{motion4}
 &&   (\omega - \mu ) F_{ml'}^{\rm c}(\omega)  \\
 &&   ~~~~ =  2J\sum_{\delta} \sigma_{m}\sigma_{m+\delta}  [ S' - \langle \sigma_{m+\delta}\sigma_{m+\delta}a_{m+\delta}^{\dagger}a_{m+\delta}               \rangle^{\rm c} \nonumber  \\
  &&   ~~~~   {}  + \langle \sigma_{m}\sigma_{m+\delta}a_{m+\delta}^{\dagger}b_{m}^{\dagger} \rangle^{\rm c}] (G_{m+\delta,l'}^{\rm c}(\omega)-F_{ml'}^{\rm c}(\omega)) \ .  \nonumber 
\end{eqnarray}
We approximate terms in the bracket [$\cdots$] in eqs. (\ref{motion3}) and (\ref{motion4}) with those averaged over impurity configurations, i.e.,
\begin{equation}
     C(\delta) \equiv S'
      - \langle \sigma_{l+\delta}\sigma_{l+\delta}b_{l+\delta}^{\dagger}b_{l+\delta} \rangle +      \langle \sigma_{l}\sigma_{l+\delta}a_{l}b_{l+\delta} \rangle \ .
\end{equation}
The correlation function $C(\delta)$ must be obtained self-consistently.

In much the same way as the above spin-wave theory, we introduce the coherent exchange integral $J_{\rm p}(\omega)$ and get
\begin{eqnarray}
    \label{matrix1}
&&  \hspace*{1.2cm} \Gamma^{-1} g^{\rm c} = \Lambda^{\rm c} + V^{\rm c} g^{\rm c} \ , \\
  \Gamma^{-1} &\equiv & 
     \left(
       \begin{array}{cc}
          (\Gamma^{-1})_{li} & (\Gamma^{-1})_{lj} \\
          (\Gamma^{-1})_{mi} & (\Gamma^{-1})_{mj}
       \end{array}
     \right)  \nonumber  \\ 
   &=& \left(
       \begin{array}{c}
       [\omega+\mu-2J_{\rm p}(\omega) \sum_{\delta}C(\delta)] \delta_{li}   \\ 
       -2J_{\rm p}(\omega) \sum_{\delta}C(\delta) \delta_{m+\delta ,i} 
       \end{array}
     \right.   \nonumber   \\
&&  \hspace*{1.5cm} \left.
       \begin{array}{c}
          2J_{\rm p}(\omega) \sum_{\delta}C(\delta) \delta_{l+\delta ,j}    \\ 
         ~[\omega-\mu+2J_{\rm p}(\omega) \sum_{\delta}C(\delta)] \delta_{mj} 
       \end{array}
     \right)   \  ,   \nonumber   \\   
g^{\rm c} &\equiv &
     \left(
       \begin{array}{c}
            G_{ll'}^{\rm c}(\omega) \\ F_{ml'}^{\rm c}(\omega) 
       \end{array}
     \right)  \ , \hspace{0.5cm} \Lambda^{\rm c} \equiv 
     \left(
       \begin{array}{c}
             \sigma_{l}\delta_{ll'} \\ 0
       \end{array}
     \right) \ , \nonumber \\
   V^{\rm c} &\equiv &
     \left(
       \begin{array}{cc}
          V_{li}^{\rm c} & V_{lj}^{\rm c} \\
          V_{mi}^{\rm c} & V_{mj}^{\rm c}
       \end{array}
     \right)  \nonumber \\ 
  &=& 
     \left(
       \begin{array}{c}
          2\sum_{\delta}[J\sigma_{l}\sigma_{l+\delta}-J_{\rm p}(\omega)]C(\delta)\delta_{li}     \\
          2\sum_{\delta}[J\sigma_{m}\sigma_{m+\delta}-J_{\rm p}(\omega)]C(\delta)\delta_{m+\delta ,i} 
       \end{array}
    \right.  \nonumber  \\
&& \hspace*{0.8cm} \left.
       \begin{array}{c}
         -2\sum_{\delta}[J\sigma_{l}\sigma_{l+\delta}-J_{\rm p}(\omega)]C(\delta)\delta_{l+\delta ,j}     \\
         -2\sum_{\delta}[J\sigma_{m}\sigma_{m+\delta}-J_{\rm p}(\omega)]C(\delta)\delta_{mj} 
       \end{array}
    \right) \ . \nonumber 
\end{eqnarray}
Using the coherent potential approximation with the two-site approximation, we obtain Green's functions and $J_{\rm p}(\omega)$, i.e.,
\begin{eqnarray}
  \label{gmsw}
 && \langle G_{ll'}^{\rm c}(\omega)\rangle_{\rm av} = \frac{2}{N}\sum_{\mibs k}
       \frac{\omega \langle \sigma_{l'}\rangle_{\rm av} + \lambda}
{\omega^{2}-\lambda^{2}(1-\eta^{2}\gamma_{\mibs k}^{2})}{\rm e}^{{\rm i}{\mibs k}\cdot({\mibs l}-{\mibs l}')} , \\
  \label{fmsw}
 && \langle F_{ml'}^{\rm c}(\omega)\rangle_{\rm av} = \frac{2}{N}\sum_{\mibs k}
       \frac{\lambda \eta \gamma_{\mibs k}}
{\omega^{2}-\lambda^{2}(1-\eta^{2}\gamma_{\mibs k}^{2})}{\rm e}^{{\rm i}{\mibs k}\cdot({\mibs m}-{\mibs l}')} , \hspace{24pt}
\end{eqnarray}
where
\begin{equation}
   \lambda \equiv 2J_{\rm p}(\omega)zC(\delta)-\mu \ , \hspace{0.5cm}
    \eta \equiv \frac{2J_{\rm p}(\omega)zC(\delta)}{\lambda} \ ,
\end{equation}
and
\begin{equation}
  \label{jpmsw}
   \frac{J_{\rm p}(\omega)}{J} = 1-\frac{2x}{1+\Gamma_{\rm p}(\omega)} \ ,
\end{equation}
with
\begin{equation}
  \label{jpfom}
   \Gamma_{\rm p}(\omega) \equiv \frac{2}{N}\sum_{\mibs k}\frac{\omega^{2}-\lambda^{2}(1-\eta)} {\omega^{2}-\lambda^{2}(1-\eta^{2}\gamma_{\mibs k}^{2})} \ .
\end{equation}
For simplicity, we consider the case that $J_{\rm p}(\omega)$ is real. For cases with the spin gap, $J_{\rm p}(\omega)$ is real for all energies within the gap. If the spin gap vanishes, then the dependence of $J_{\rm p}(\omega)$ on $x$ is required to tend to that of the spin-wave theory. Therefore, we put $\omega = \lambda \sqrt{1-\eta} = \sqrt{\lambda (-\mu)}$. Then from eqs. (\ref{jpmsw}) and (\ref{jpfom}) we obtain
\begin{equation}
    \label{jp2x}   
     \frac{J_{\rm p}(\sqrt{\lambda (-\mu)})}{J} = 1-2x \ .
\end{equation}
  
From the spectral relation (\ref{spec}) and eqs. (\ref{gmsw}), (\ref{fmsw}) and (\ref{jp2x}) correlation functions are obtained:
\begin{eqnarray}
    \label{abcorr}
   &&    \langle \sigma_{l}\sigma_{l'}a_{l}^{\dagger}a_{l'} \rangle 
         = f(l-l')-\frac{1}{2}\delta_{ll'} \langle \sigma_{l'} \rangle_{\rm av} \ , \\
    \label{abcorr2}
   &&    \langle \sigma_{l}\sigma_{m}a_{l}^{\dagger}b_{m}^{\dagger} \rangle = g(l-m) \ ,
\end{eqnarray}
where
\begin{eqnarray}
    \label{fgdef}
 &&   f(l-l') \equiv \frac{2}{N}\sum_{\mibs k} \frac{1}{\sqrt{1-\eta^{2}\gamma_{\mibs k}^{2}}}
      \hspace{3cm} \nonumber \\
 &&  \hspace*{48pt} {} \times \frac{1}{2}{\rm coth}(\frac{\beta \lambda}{2}\sqrt{1-\eta^{2}\gamma_{\mibs k}^{2}})
             {\rm e}^{-{\rm i}{\mibs k}\cdot({\mibs l}-{\mibs l}')} \ , \\
    \label{fgdef2}
 &&   g(l-m) \equiv \frac{2}{N}\sum_{\mibs k} \frac{\eta \gamma_{\mibs k}}
             {\sqrt{1-\eta^{2}\gamma_{\mibs k}^{2}}} \hspace{3cm} \nonumber \\
 &&  \hspace*{48pt} {} \times \frac{1}{2}{\rm coth}(\frac{\beta \lambda}{2}\sqrt{1-\eta^{2}\gamma_{\mibs k}^{2}})
             {\rm e}^{-{\rm i}{\mibs k}\cdot({\mibs l}-{\mibs m})} \ . 
\end{eqnarray}
In the modified spin-wave theory we put the spontaneous sublattice magnetization zero.  Since nonmagnetic impurities with a completely random configuration cannot change rotational symmetry of the system, we can put the magnetization zero for impure systems:
\begin{equation}
  \label{zeromag}
   0 = \langle \sigma_{l}S_{l}^{\rm z} \rangle = (S+\frac{1}{2})\langle \sigma_{l} \rangle_{\rm av} -               f(0) \ .  
\end{equation}
From eqs. (\ref{abcorr}), (\ref{abcorr2}) and (\ref{zeromag}) correlation functions of spin operators and $C(\delta)$ are written by
\begin{eqnarray}
   \label{spincorr2} 
 &&  \langle \sigma_{l}\sigma_{l'}{\mib S}_{l}\cdot {\mib S}_{l'} \rangle = [f(l-l')]^{2} 
        \hspace{24pt} {\rm for} \ l' \neq l  \ , \\
   \label{spincorr2g}
 &&  \langle \sigma_{l}\sigma_{m}{\mib S}_{l}\cdot {\mib S}_{m} \rangle = -[g(l-m)]^{2} \ , \\
   \label{cdel}
 &&  C(\delta) = g(\delta) \ .
\end{eqnarray}
The on-site correlation function at $l'=l$ is $\langle \sigma_{l} {\mib S}_{l} \cdot {\mib S}_{l} \rangle = S(S+1)\langle \sigma_{l} \rangle_{\rm av} = S(S+1)(1-x)$.

At $T=0$ from eqs. (\ref{fgdef}) and (\ref{zeromag}) we have
\begin{equation}
   \label{t0eta}
     (S+\frac{1}{2})(1-x) = 
          \frac{2}{N}\sum_{\mibs k}\frac{1}{2\sqrt{1-\eta^{2}\gamma_{\mibs k}^{2}}} \ .
\end{equation}
We take the limit $N\to \infty$ and rewrite  the sum into the integral for the right-hand side of eq. (\ref{t0eta}). Although the sum is divergent as $\eta \to 1$, the integral is not divergent. This contradiction can be avoided by introducing Bose-Einstein condensation.[16] Separating the divergent term at ${\mib k}=(0,0)$ from the sum, we have 
\begin{equation}
   \label{bose}
   (S+\frac{1}{2})(1-x) = \frac{1}{N\sqrt{1-\eta^{2}}}
        + \frac{1}{2}\int \frac{{\rm d}{\mib k}}{(2\pi )^{2}}\frac{1}{\sqrt{1-\gamma_{\mibs k}^{2}}} \ .
\end{equation}
The mean-square root of the staggered magnetization $m$ is defined by 
\begin{eqnarray}
  \label{order}
  m^{2} &\equiv& \langle [\frac{1}{N}\sum_{n}(-1)^{n}\sigma_{n}{\mib S}_{n}]^{2} \rangle \nonumber \\
        &=&     \frac{1}{N}\sum_{n}(-1)^{n}
               \langle \sigma_{0}\sigma_{n}{\mib S}_{0}\cdot {\mib S}_{n} \rangle \ .
\end{eqnarray}
This is the long range order (LRO) parameter in the thermodynamic limit. From eqs. (\ref{fgdef}), (\ref{fgdef2}), (\ref{spincorr2}), (\ref{spincorr2g}) and (\ref{order}) LRO parameter is given by 
\begin{eqnarray}
 &&  m^{2}=\frac{1}{2N^{2}}\sum_{\mibs k}\frac{1+\eta^{2}\gamma_{\mibs k}^{2}}{1-\eta^{2}\gamma_{\mibs k}^{2}}
      {\rm coth}^{2}(\frac{\beta \lambda}{2}\sqrt{1-\eta^{2}\gamma_{\mibs k}^{2}}) \nonumber \\
 && \hspace*{4cm} {} - \frac{1}{4N}(1-x) \ .
\end{eqnarray}
For $J_{\rm p}\neq 0$ we take the limit $T\to 0$ on finite lattice. From $\eta \neq 1$ we have
\begin{equation}
   \label{m2}
   m^{2}=\frac{1}{2N^{2}}\sum_{\mibs k}\frac{1+\eta^{2}\gamma_{\mibs k}^{2}}{1-\eta^{2}\gamma_{\mibs k}^{2}}
       - \frac{1}{4N}(1-x) \ .
\end{equation}
In the limit $N\to \infty$, integrating eq. (\ref{m2}) except ${\mib k}=(0,0)$ and neglecting $x/4N$, we obtain
\begin{equation}
   \label{LRO}
   m^{2} = \frac{1}{N^{2}}\frac{1}{1-\eta^{2}} \ .  
\end{equation}
From eqs. (\ref{bose}) and (\ref{LRO}) LRO parameter at $T=0$ is written by 
\begin{eqnarray}
     m &=& (S+\frac{1}{2})(1-x)
  - \frac{1}{2}\int \frac{{\rm d}{\mib k}}{(2\pi )^{2}}\frac{1}{\sqrt{1-\gamma_{\mibs k}^{2}}} \nonumber \\
    &=& (S+\frac{1}{2})(1-x) - 0.696602 \ .
\end{eqnarray}
Thus $x_{\rm c}$ and $x_{\rm p}$ are agreement with those of the spin-wave theory. Our modified spin-wave theory has the solution with $\eta \neq 1$ in $x_{\rm c}<x<x_{\rm p}$ from eq. (\ref{t0eta}). The state in this concentration range is the disordered state with the spin gap. We conclude that ${\rm La}_{2}{\rm Cu}_{1-x}{\rm Mg}_{x}{\rm O}_{4}$ has the disordered state with the spin gap in $x_{\rm c}<x<x_{\rm p}$, but ${\rm K}_{2}{\rm Mn}_{1-x}{\rm Mg}_{x}{\rm F}_{4}$ is in the N\'{e}el state at all concentrations $x<x_{\rm p}$. This disordered state is presumably due to the construction of the singlet dimer bonds.

\section{Summary and Discussion}

In the present paper we have applied the spin-wave theory and the coherent potential approximation (CPA) to the impure 2D Heisenberg antiferromagnet with spin $S$. As magnetic ions of spin $S$ are substituted by nonmagnetic impurities, both the number of magnetic ions and interacting spin pairs are decreased. We take these two effects into account by substituting $S(1-x)$ for $S$ and using CPA to the exchange integral $J\sigma_{i} \sigma_{j}$. At $T=0$ the critical concentration $x_{\rm c}$ is 0.303 for $S=1/2$ and 0.500 for $S\geq 1$. On the other hand, the percolation threshold $x_{\rm p}$ is 0.500 for all spin $S$'s. Thus for $S=1/2$ the state, with the infinite connecting percolation path and without long range order, exists in $x_{\rm c}<x<x_{\rm p}$. For $S\geq 1$ the long range order vanishes at the percolation threshold. The decrease in the effective spin leads the difference between $x_{\rm c}$ and $x_{\rm p}$ for $S=1/2$. To investigate the state in $x_{\rm c}<x<x_{\rm p}$, we have studied by the modified spin-wave theory. As a result, this state is the disordered state characterized by the finite spin gap and the exponentially decaying correlation function. This disordered state is presumably due to the construction of  the singlet dimer bonds.

Thus our results qualitatively explain the experimental results: the critical impurity concentration $x_{\rm c}$ for the N\'{e}el state of ${\rm La}_{2}{\rm Cu}_{1-x}{\rm Mg}_{x}{\rm O}_{4}~(S=1/2)$ is smaller than the percolation threshold $x_{\rm  p}=0.41$, on the other hand, $x_{\rm c}$ of ${\rm K}_{2}{\rm Mn}_{1-x}{\rm Mg}_{x}{\rm F}_{4}~(S=5/2)$ is equal to $x_{\rm p}$. ${\rm La}_{2}{\rm Cu}_{1-x}{\rm Mg}_{x}{\rm O}_{4}$ would be the disordered state in $x_{\rm c}<x<x_{\rm p}$. We expect to experimentally observe the spin gap in low temperatures for ${\rm La}_{2}{\rm Cu}_{1-x}{\rm Mg}_{x}{\rm O}_{4}$ in $x_{\rm c}<x<x_{\rm p}$ or for other materials with $S=1/2$.
 
\section*{Acknowledgment}

We would like to thank Dr. Y. Fukumoto for valuable discussions.


\end{document}